\documentstyle[preprint,prl,aps,epsf]{revtex}
\begin{document}
\draft
\title{Upper Limit on the Decay $K^{+}\rightarrow e^+ \nu \mu^+\mu^{-}$}


\author{S. Adler, M.S. Atiya, I-H. Chiang, J.S. Frank, J.S. Haggerty,
  T.F. Kycia, K.K. Li, \\ L.S. Littenberg, 
  A. Sambamurti\cite{AKS}, A. Stevens, R.C. Strand, and C. Witzig}
\address{
  Brookhaven National Laboratory, Upton, New York 11973
}

\author{W.C. Louis}
\address{
  Medium Energy Physics Division, Los Alamos National Laboratory, \\
Los Alamos, New Mexico 87545
}

\author{D.S. Akerib\cite{DSA}, M. Ardebili\cite{MA}, M. Convery\cite{MC}, 
M.M. Ito\cite{MMI}, 
D.R. Marlow, R. McPherson\cite{RMcP}, \\ 
P.D. Meyers, M.A. Selen\cite{MAS}, F.C. Shoemaker, and A.J.S. Smith}
\address{
  Joseph Henry Laboratories, Princeton University,
Princeton, New Jersey 08544
}

\author{E.W. Blackmore, D.A. Bryman, L. Felawka,
  A. Konaka, 
  Y. Kuno\cite{YK}, J.A. Macdonald, \\ T. Numao, 
P. Padley\cite{PP},
  J.-M. Poutissou,
  R. Poutissou, J. Roy\cite{JR}, and A.S. Turcot\cite{AT}}
\address{
  TRIUMF, Vancouver, British Columbia, Canada, V6T 2A3
}

\author{P. Kitching, T. Nakano\cite{TN}, M. Rozon\cite{MR}, and
  R. Soluk}
\address{
 Center for Subatomic Research, University of Alberta, Edmonton, Alberta, 
Canada, T6G 2N5
}


\date{\today}
\maketitle

\begin{abstract}
An upper limit on the branching ratio for the decay 
$K^{+}\rightarrow e^+\nu \mu^+\mu^{-}$ is set at $5.0 \times 10^{-7}$ at
90\% confidence level,  consistent with predictions from chiral perturbation theory.
\end{abstract}


\newpage
We report here an upper limit on the rare kaon decay
$K^{+}\rightarrow e^+\nu \mu^+\mu^{-}$. This decay together with
the related semileptonic decays $K^{+}\rightarrow l^{+} \nu l'^+l'^{-}$, where
$l$ and $l'$ stand for electron or muon, is of interest for testing the
Standard Model in next-to-leading order in the chiral expansion
without any further assumptions. For $K^{+}\rightarrow e^+\nu \mu^+\mu^{-}$ 
the calculation yields a branching ratio of $1.1 \times 10^{-8}$ \cite{bi1}.
This decay is a particularly good test of the chiral expansion because the
structure dependent terms dominate over the inner bremsstrahlung term due to 
helicity suppression.
The decay  $K^{+}\rightarrow e^+\nu \mu^+\mu^{-}$ has not been seen and no
previous limit has been reported, in contrast to the
decays $K^{+}\rightarrow e^+ \nu e^+e^{-}$ and  $K^{+}\rightarrow \mu^+ \nu e^+e^{-}$,
which have been observed \cite{eee} and $K^{+}\rightarrow \mu^+ \nu \mu^+\mu^{-}$ for which
an upper limit has been reported \cite{mmm}.

\begin{figure}
\leavevmode
\hfil\epsfysize=3.0in\epsfbox{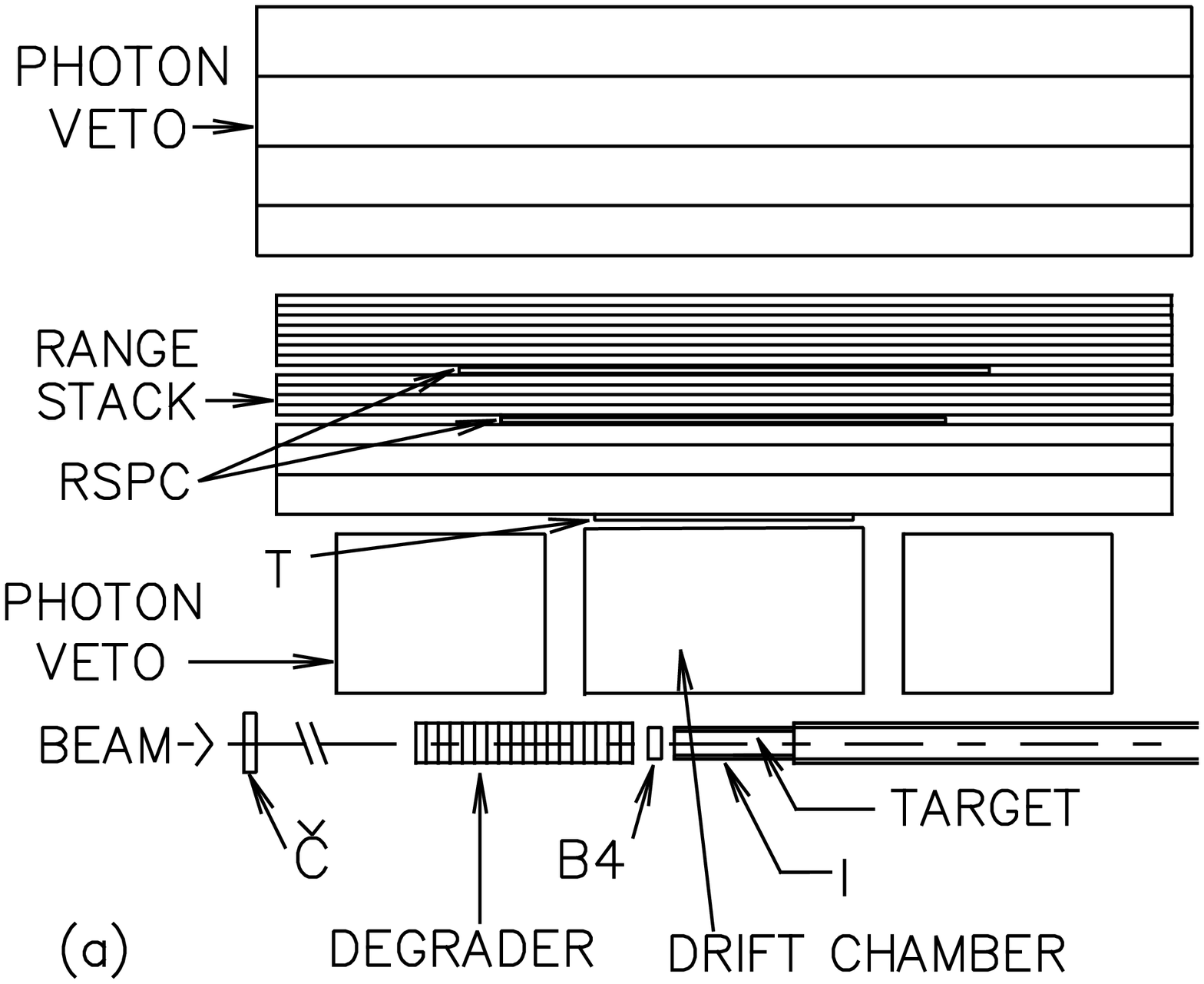}\hfil
\hfil\epsfysize=3.0in\epsfbox{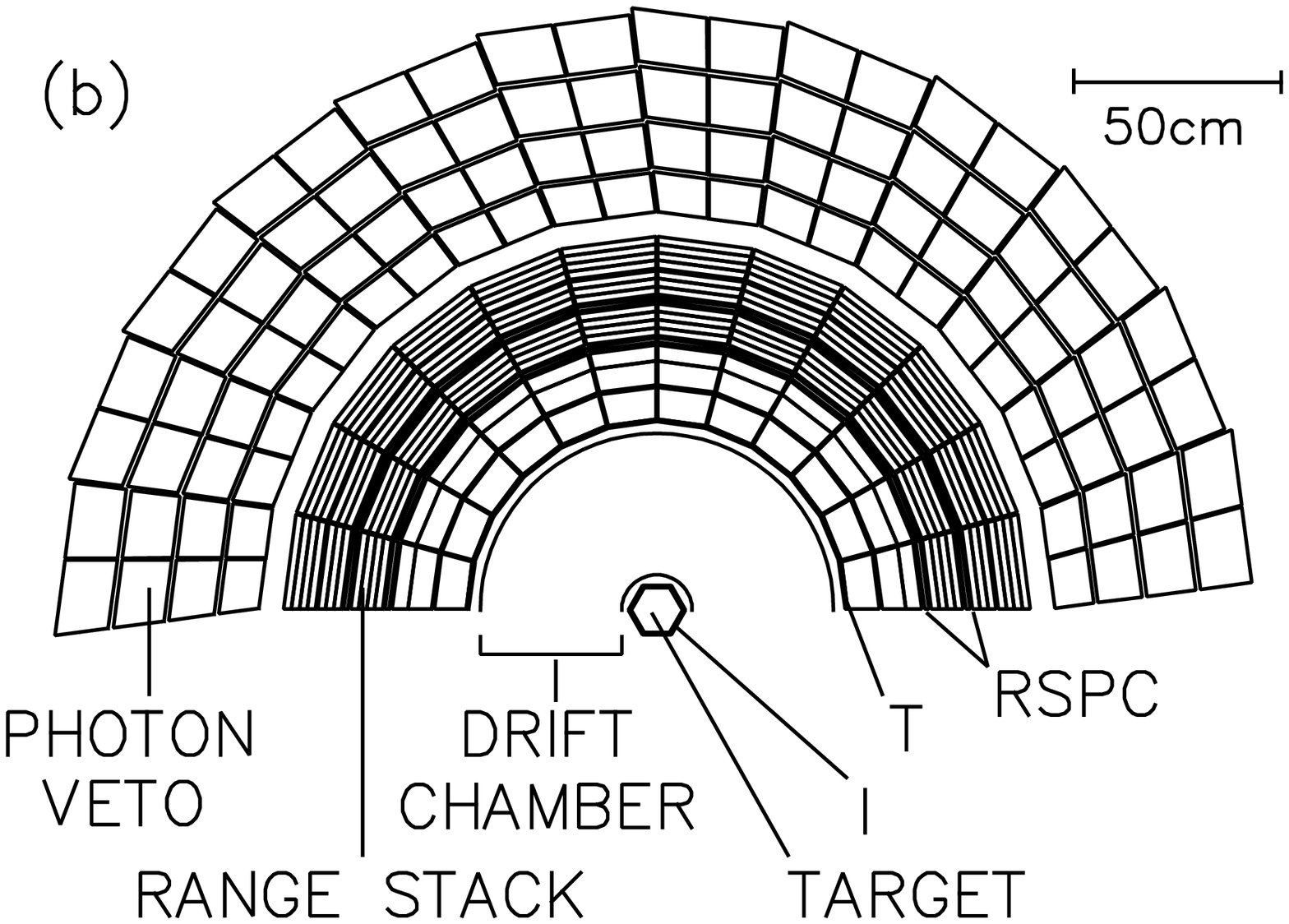}\hfil
\caption{Schematic (a) side and (b) end views showing the upper
half of the E787 detector. \v C: beam \v Cerenkov counter, B4: beam
hodoscope, I and T: trigger scintillators, RSPC: multiwire proportional
chambers.}
\label{fig1}
\end{figure}

The experiment described  here used the E787 apparatus \cite{e787nim}
at the Brookhaven National Laboratory Alternating Gradient Synchrotron. The data
were taken between 1989 and 1991. Kaons of 800 MeV$/c$ momentum were tagged by a 
\v Cerenkov detector and subsequently stopped in a scintillating
fiber target located in the center of the detector (see Fig. \ref{fig1}).  
Six trigger counters
surrounding the target defined the decay volume of the kaons.
Charged decay products of the kaon with sufficiently high momentum could
leave the target and have their momentum
determined in a cylindrical tracking chamber. A
range stack of scintillators divided radially into 15 layers and 
azimuthally into 24 sectors surrounded the tracking chamber.
The innermost layer, with a thickness of 0.6 cm,  served as trigger counter;
the next three layers were 7.6, 5.7 and 3.8 cm thick followed by 11 1.9-cm
layers.
Phototubes on both ends of the range stack were read out 
with 500-MHz transient digitizers (TD) \cite{e787td}.
A momentum of at least 65 MeV$/c$ was needed for a particle to reach the range stack. 
Two Pb-scintillator sandwich photon veto systems completed the detector: a ``barrel veto'' 
of 14 radiation lengths in the radial direction and two ``end cap vetoes'' of
12 radiation lengths on the upstream and downstream ends of the detector.
The entire apparatus was in a 1-T solenoidal magnetic field.
The beam delivered a 1.2-s spill every 3.0 s. Approximately
$3\times 10^5$ kaons stopped in the target per spill, and the dead-time-corrected 
total exposure was $3 \times 10^{11}$ stopped kaons.

The trigger was optimized for the search for $K^+ \rightarrow \pi^+ \mu^+ \mu^-$ \cite{e787pmm}.
It rejected kaon decays in flight by imposing a delay of at least 1.5 ns 
between the incident $K^+$ and its decay particles
and required two or three charged tracks in the range stack, with none penetrating
deeper than 14 cm at $90^{\circ}$.
Events with more than 5 MeV in the barrel veto, 10 MeV in the end cap or 5 MeV in the outer
regions of the range stack were rejected by the trigger.

The search for $K^{+}\!\rightarrow\!e^{+}\nu\mu^{+}\mu^{-}$ used the same data
set and preliminary event selection  as Ref. \cite{e787pmm}. First it was demanded that
three charged tracks, each with momentum less than 172 MeV$/c$, were found in the drift chamber, 
pointing  back to the kaon stopping region as determined by the target reconstruction software. 
Photon veto requirements beyond those in the trigger were applied to reject further the copious
kaon decays with  $\pi^0$s.
Good accceptance (80\%) for showering electrons was retained by ignoring regions around the
charged tracks in these tighter photon vetoes.
$14\,200$ out of $6 \times 10^6$ recorded events passed these requirements.

The major backgrounds at this stage of the analysis were
$K^{+}\!\rightarrow\! \pi^{+}\pi^{-} e^{+} \nu$  ($K_{e4}$) and
$K^+\!\rightarrow\!\pi^0\mu^+\nu$ ($K_{\mu3}$) where one of the photons from the 
$\pi^0$ underwent internal or external conversion in the target to produce
the additional two charged tracks in the drift chamber.
The $K_{\mu3}$ background was
strongly reduced by rejecting events where a pair of oppositely charged
drift chamber tracks was found to have a low invariant mass ($m_{ee}$ $<$ 20 MeV$/c^2$),
assuming electron masses for the particles \cite{rejection_comment}.
The next step of the analysis consisted of using three sets of particle identification
methods to separate signal from background.
First, a maximum likelihood  analysis was performed combining 
$dE/dx$ information from the drift chamber, $dE/dx$ in the first layer of
the range stack, and time of flight from the trigger counter to the range stack.
For particles not reaching the range stack the
maximum likelihood analysis was replaced by a simple $dE/dx$ cut on the drift chamber information.
The likelihood function was normalized using reference data
samples and set to identify 90\% of the electrons in the momentum region of interest.
Second, the negative charged track was required to enter the range stack and
the TDs were used to search for a subsequent decay electron.
This distinguishes  $\mu^-$ from electrons and  $\pi^-$ (the latter 
being predominantly absorbed by a carbon nucleus in scintillator \cite{pi-C}). The efficiency
of this electron search was found to be about 67\%.
Third, two requirements on kinematic quantities in the range stack were demanded.
The masses of the two charged particles in the range stack were calculated
from the momentum measured in the tracking chamber and the kinetic energy
measured in the range stack and required to be
between 60 MeV$/c^2$ and 150 MeV$/c^2$. (The r.m.s. resolution on the mass was 
10-20 MeV$/c^2$ for pions and muons.) 
In order to obtain meaningful masses, it was demanded that the two range stack
tracks did not have any counters in common. In addition, it was required
that both tracks reached the third layer of the range stack in order to further
select muons over pions.

To summarize the particle identification
requirements: Tracks entering the range stack had to be  consistent
with an oppositely charged muon pair hypothesis 
and the third track, which was not required to enter the range stack,
had to have an electron-like particle identification signature.

\begin{figure}
\leavevmode
\hfil\epsfysize=3.0in\epsfbox{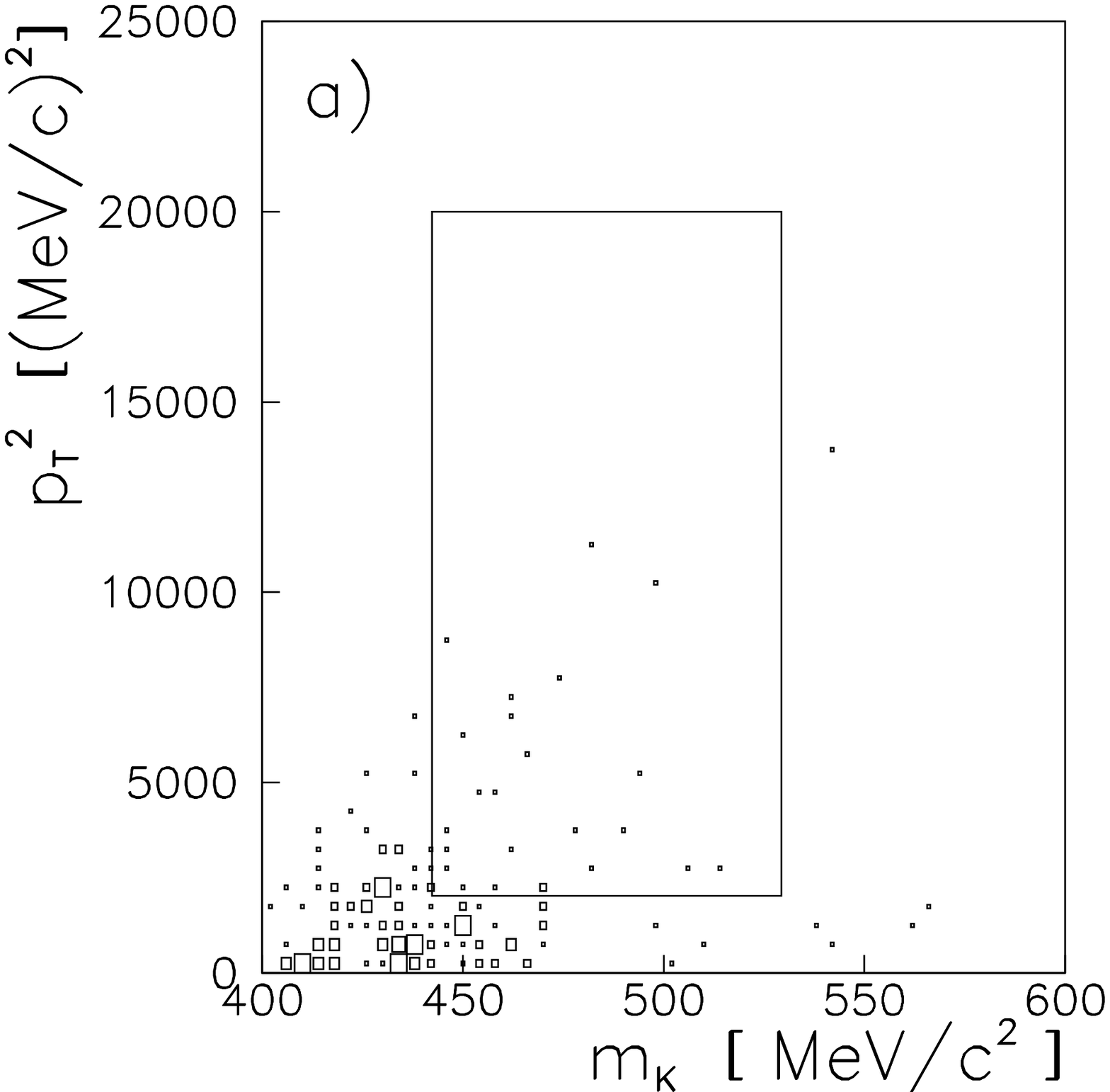}\hfil
\hfil\epsfysize=3.0in\epsfbox{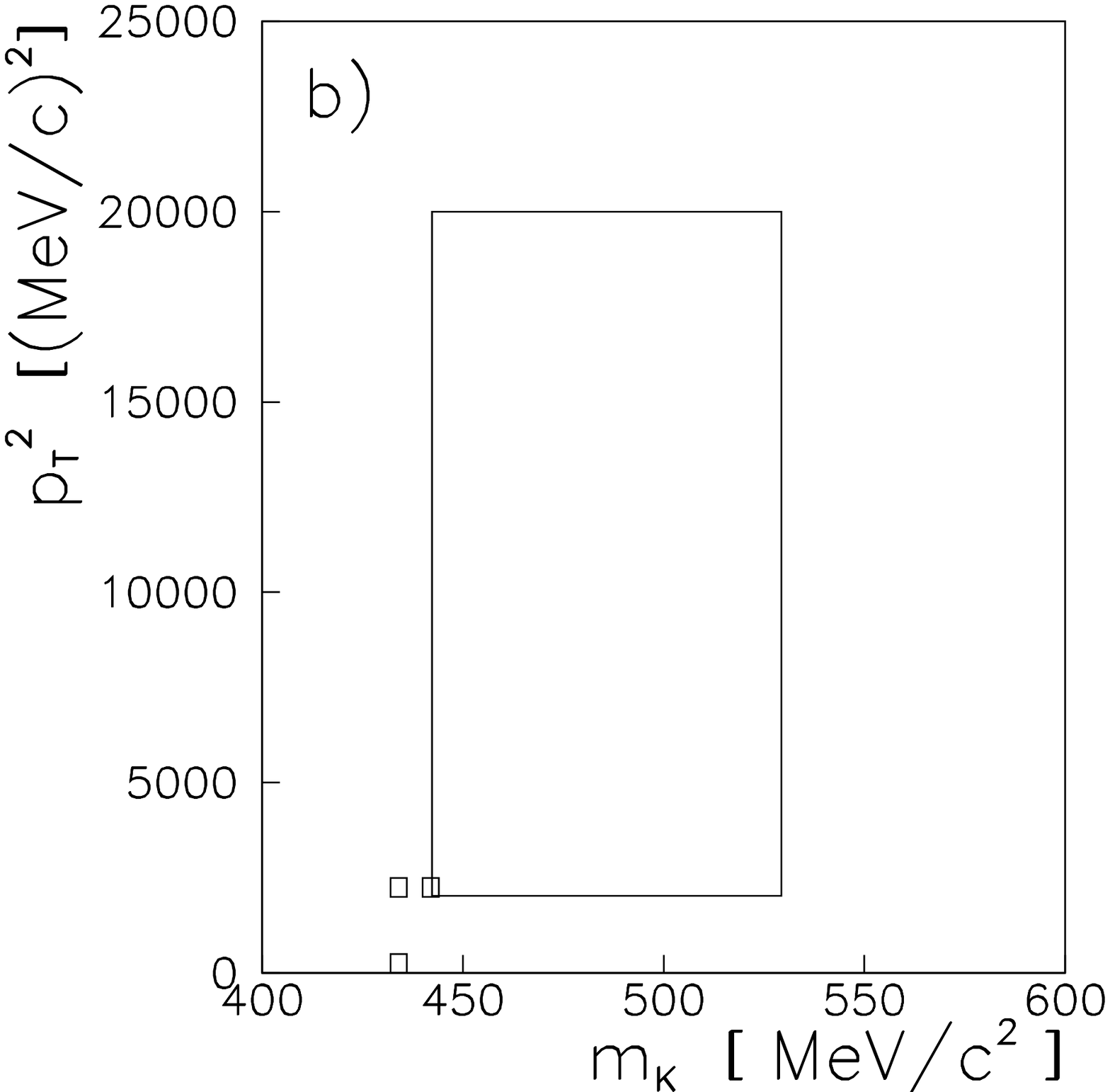}\hfil
\hfil\epsfysize=3.0in\epsfbox{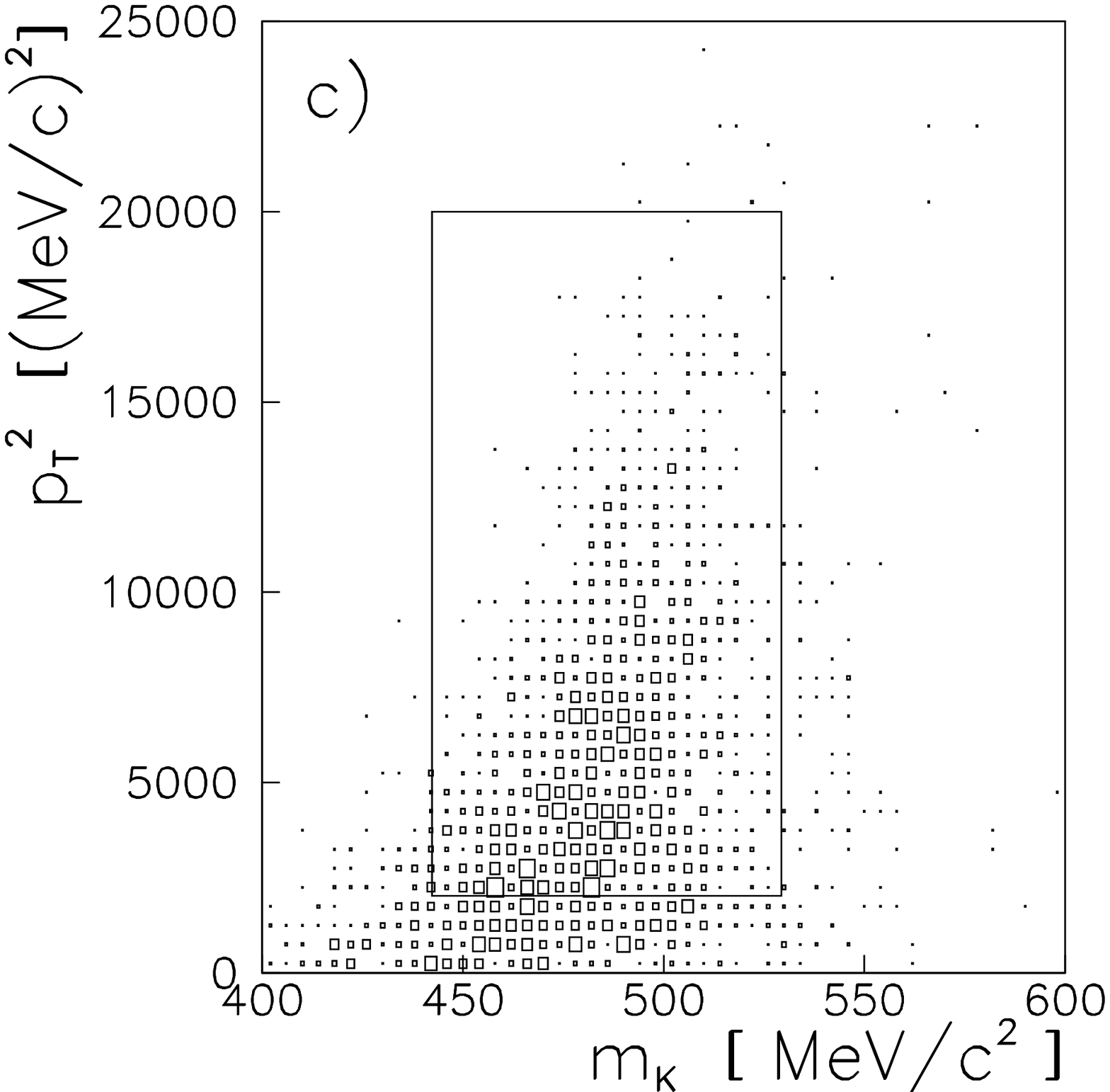}\hfil
\hfil\epsfysize=3.0in\epsfbox{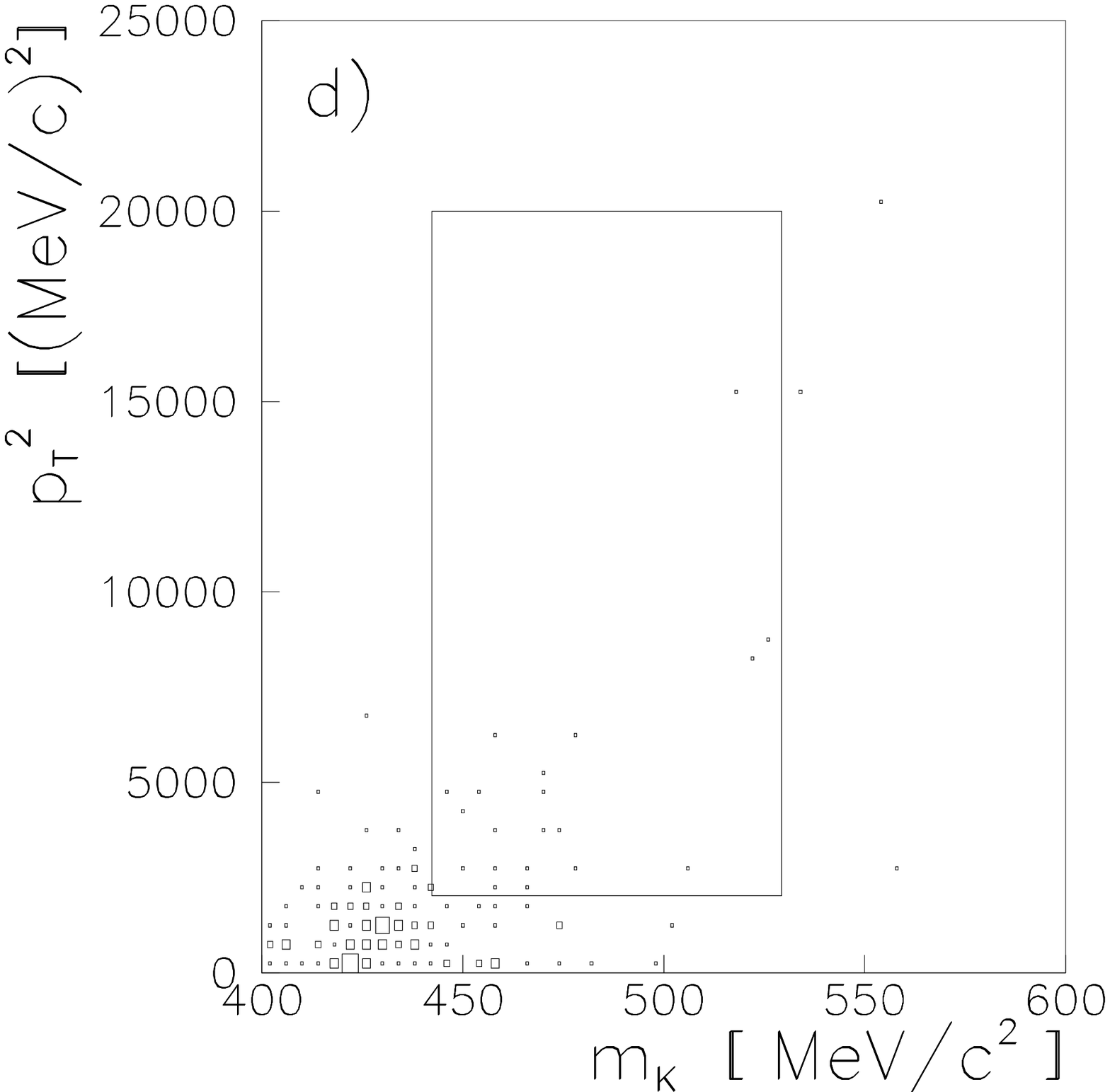}\hfil
\caption{Sum of the transverse momentum squared of the three charged particles
at the kaon vertex vs.~reconstructed
kaon mass a) at an earlier stage of the analysis, b) after all rejection criteria have been
applied, c) for simulated $K^+ \rightarrow e^+ \nu \mu^+ \mu^-$ events, and d) for the simulated
background of $K^+ \rightarrow \pi^+ \pi^- e^+ \nu$ and $K^+ \rightarrow \pi^0 \mu^+ \nu$
at the same stage of the analysis as figure 2a).}
\label{fig2}
\end{figure}

In the final stage of the analysis the three charged tracks were extrapolated back
to the decay vertex, correcting their momenta for energy loss in the target. The missing
momentum was assigned to a neutrino, and the invariant mass of the full
four-track event was calculated. The signal region
was defined by demanding the invariant mass be less than two standard
deviations away from the kaon mass ($442.3 < m_K <  529.3$  MeV/$c^2$) and the sum of the transverse
momenta of the charged particles be larger than 45 MeV$/c$. $K_{e4}$ events are 
primarily confined to the region 
below 45 MeV$/c$. The acceptance of this signal region was 61\%, primarily due to 
the requirement on the transverse momentum.
No event was found in the signal region and three were found outside, 
of which one was very close to the edge of the signal region (see Fig. \ref{fig2}).

The background was estimated to be 0.32 events in the signal region. The main 
contribution was identified to be $K_{e4}$ decays
with a particle identification error (0.23 events) and  $K_{e4}$ 
decays in which the $\pi^-$ decayed in flight before reaching the range stack (0.02 events).
The background due to $K_{\mu3}$ was estimated to be 0.07 events.

As a cross check of the background estimate, the  requirement that both range stack
tracks extend to the third layer was removed.
Seven events were found in reasonable agreement with the expected background of 4.4 events.

A normalization sample of $K^+\!\rightarrow\!\mu^+\nu$ events, taken 
simultaneously with the data, was used to determine the number of stopped
kaons. Data were used to determine accidental losses and the acceptance of the 
rejection criteria based on 
timing quantities and particle identification.
Monte Carlo simulation using the decay matrix element of Ref. \cite{bi1} 
was used to  calculate the acceptance of the
trigger and the kinematical cuts.
The full acceptance was determined to vary between $(1.4 \pm 0.2) \times  10^{-5}$
and $(1.6 \pm 0.3)  \times  10^{-5}$ depending on the year in which the data were taken. Table 
\ref{acc_table} lists the main acceptance factors for one year (1990). 

As a check of the acceptance calculation, the branching ratio for $K_{e4}$ decay
was measured to be $(3.51 \pm 0.57) \times 10^{-5}$ \cite{comment_error}.
The world average for this decay is  $(3.91 \pm 0.17) \times 10^{-5}$ \cite{ke4_br}.

Based on zero observed events and a single
event sensitivity of $2.2 \times 10^{-7}$, we obtain a  90\% confidence level upper limit of 
$B(K^{+}\!\rightarrow\! e^+ \nu \mu^+ \mu^-) < 5.0 \times 10^{-7}$, consistent with the
prediction from chiral perturbation theory of $1.1 \times 10^{-8}$ \cite{bi1}.


\acknowledgments

We gratefully acknowledge the dedicated efforts of the technical staffs
supporting this experiment and of the Brookhaven AGS Department.
This research was supported in part by the U.S. Department of Energy
under contracts DE-AC02-76CH00016, W-7405-ENG-36, and 
grant DE-FG02-91ER40671
and by the Natural Sciences and Engineering Research Council and the
National Research Council of Canada.



\begin{table}[hbt]
\begin{center}
\begin{tabular}{l r }
\hline
Acceptance factors      &      \\
\hline
Trigger  & 0.010 \\
Photon veto acceptance and loss due to accidentals & 0.64 \\
3 tracks in the drift chamber & 0.24 \\
Other reconstruction efficiencies  & 0.49 \\
Particle identification & 0.30 \\
Other kinematic cuts & 0.11  \\
Signal region ($442.3 \leq m_K \leq 529.3$ MeV/$c^2$ and $p_T^2 \leq 45^2$ MeV$/c$) & 0.61 \\
\hline
Total acceptance & $1.6 \times 10^{-5}$ \\
\hline
\end{tabular}
{\caption
{\label{acc_table}
Acceptance factors used in the search for $K \rightarrow e^+ \nu \mu^+ \mu^-$ for 1990 (the other
years are comparable). ``3 charged tracks in the drift chamber'' is the requirement that all
three tracks leave the target and are successfully reconstructed in the drift chamber.
``Other reconstruction efficiencies'' includes full reconstruction in the 
target and cuts in the beam counters. ``Other kinematic cuts'' includes the requirements that the tracks in 
the range stack did not share any counters and  reach the third layer.}}
\end{center}
\end{table}

\begin{references}
\bibitem{rejection_comment} The rejection factor for this cut on the 
$K_{\mu3}$ background was 7 at an acceptance loss of 14\%.
\bibitem{comment_error} The statistical and systematic
error have been added in quadrature. The error is dominated by the systematic
error whose biggest contribution is the uncertainty of acceptance loss due to $\pi^-$
absorption effects.
\end{references}
\end{document}